\documentclass[aps,prd,groupedaddress,showpacs,nofootinbib
]{revtex4}
\usepackage{graphicx,amssymb,amsmath,bm}

\newcommand{\be}{\begin{equation}}
\newcommand{\ee}{\end{equation}}
\newcommand{\bea}{\begin{eqnarray}}
\newcommand{\eea}{\end{eqnarray}}
\newcommand{\nab}{\nabla}

\newcommand{\la}{\label}

\def\de#1/de#2{\frac{\partial {#1}}{\partial {#2}}}
 
\begin{document}

\title{Spinor fields in $f(\mathcal{Q})$-gravity}

\author{ Stefano Vignolo$^{1}$\footnote{E-mail: stefano.vignolo@unige.it}, Sante Carloni$^{1}$\footnote{E-mail: sante.carloni@unige.it}, Roberto Cianci$^{1}$\footnote{E-mail: roberto.cianci@unige.it}, Fabrizio Esposito$^{1}$\footnote{E-mail: fabrizio.esposito01@edu.unige.it} and Luca Fabbri$^{1}$\footnote{E-mail: fabbri@dime.unige.it}}

\affiliation{ $^{1}$ DIME, Universit\`{a} di Genova, Via all'Opera Pia 15, 16145 Genova, ITALY}

\begin{abstract}
We present a tetrad--affine approach to $f(\mathcal{Q})$ gravity coupled to spinor fields of spin-$\frac{1}{2}$. After deriving the field equations, we derive the conservation law of the spin density, showing that the latter ensures the vanishing of the antisymmetric part of the Einstein--like equations, just as it happens in  theories with torsion and metricity. We then focus on Bianchi type-I cosmological models proposing a general procedure to solve the corresponding field equations and providing analytical solutions in the case of gravitational Lagrangian functions of the kind $f(\mathcal{Q})=\alpha\mathcal{Q}^n$. At late time such solutions are seen to isotropize and, depending on the value of the exponent $n$, they can undergo an accelerated expansion of the spatial scale factors.   
\end{abstract}
\date{\today}
\maketitle

\section{introduction}
General Relativity (GR) has been the first theory in which one of the fundamental forces is entirely described in geometrical terms. Using the tools of differential geometry, Einstein discovered that it is possible to describe the gravitational phenomena entirely in terms of the geometrical properties of spacetime. However, as it often happens, geometry offers much more degrees of freedom than the ones necessary for physical theories. In particular, torsion and non--metricity, which are considered zero in the original version of GR, can modify and expand the types of geometry of the spacetime. Einstein himself became aware of this issue and tried to use (unsuccessfully) these additional degrees of freedom to pursue his attempt of geometrization and unification of the fundamental forces (see e.g.\cite{Goenner}). 

Indeed, it has been known for some time that torsion and non--metricity can encode the entirety of the degrees of freedom of the gravitational field so that one can consider a flat spacetime in which the gravitational field is mediated only by torsion and non--metricity. This property is well represented by the fact that it is possible to define scalar quantities that can be used to construct actions that, modulo of a boundary term, reproduce in full the equations of GR.

Another interesting aspect of these formulations of the theory of gravitation is that, because of the nature of the connection terms associated with torsion and non--metricity, one can consider non-linear Lagrangians without having the problem of obtaining higher-order terms in the field equations.

Today we know that torsion can be associated with a purely quantum property of matter, i.e., the spin. This realization led to the development of the so-called Einstein-Cartan-Sciama-Kibble (ECSK) theory \cite{S,K,h-h-k-n,Hehl:2007bn} and its several extensions \cite{German,Rubilar,CCSV,KB,BH,VFS,Fabbri:2014dxa}, which allow one to introduce spin degrees of freedom in a relativistic geometric theory. 

Very recently, there has been the proposal to connect dark components of the universe with non--metricity terms and, in particular, to non-linear $f(\mathcal{Q})$ Lagrangians written in terms of the non--metricity scalar $\mathcal{Q}$. These $f(\mathcal{Q})$ theories represent a generalization of the Symmetric Teleparallel Gravity (STG), like $f(R)$ gravity with or without torsion does for ECSK theory or GR, respectively. Indeed, assuming the same paradigm of $f(R)$ gravity, the main idea consists in relaxing the hypothesis that the gravitational Lagrangian is a linear function of the non--metricity scalar, allowing the gravitational Lagrangian to have a dependence on $\mathcal{Q}$ more general than that of STG. $f(\mathcal{Q})$ gravity has been studied in different contexts with encouraging results. For example, some cosmological applications have been analyzed in \cite{Lazkoz,Harko1,Lu,Koivisto1,Bajardi,Mandal,Barros,Frusciante}.

Naturally, any consistent and viable theory of gravitation must be able to admit couplings with every possible kind of energy-matter fields; among these, spinor fields certainly play a leading role in physics. 

In this regard, in the present paper we study $f(\mathcal{Q})$ gravity coupled to a spin-$\frac{1}{2}$ spinor field. In the authors' knowledge, this topic has not been fully explored in literature. A first step in this direction was given in \cite{Adak}, in the context of the STG. 

As $f(\mathcal{Q})$ gravity is a metric-affine theory of gravitation where the dynamical connection is not metric compatible (but torsion-free and flat), the introduction of spinor fields in $f(\mathcal{Q})$ theory requires the definition of spinor covariant derivatives induced by a general affine connection, not necessarily metric compatible. The spinor covariant derivative can be obtained by using the Fock--Ivanenko coefficients with antisymmetrized anholonomic (Lorentz) indexes \cite{Poplawski}. As a result, the covariant derivatives of Dirac matrices are no longer zero, and this is reflected in the particular form of the spin conservation law. 
 
Another peculiar aspect of the proposed approach to $f(\mathcal{Q})$ gravity coupled to spinors is the choice of the dynamical variables we used. Theories of gravity that deal with spinor fields employ tetrad fields to represent the metric tensor. In tetrad--affine theories, the additional gravitational degrees of freedom are usually incorporated by the spin (anholonomic) connection: for instance, in Einstein--Cartan--like theories, the variational derivative of the action with respect to the spin connection induces the well-known coupling between spin and torsion. Instead, in the present paper, we make a somewhat different choice in which the dynamical variables are a tetrad field and an affine connection expressed in holonomic (coordinate) basis. With this choice, the Lagrange multipliers present in the action do not enter the Einstein-like equations, and thus we obtain more manageable field equations. 

For this reason, in the following, emphasis has been given to the deduction and discussion of the gravitational field equations. Using our new formulation, we will see that the energy-momentum tensor acquires additional terms with respect to the standard Dirac tensor. These additional terms directly involve covariant derivatives of the spin density related to its conservation law. The latter has been obtained by making use of the Dirac equations derived by variations. We found that even if the final expression of the spin conservation law is formally different from that holding in Einstein--Cartan--like theories, it still ensures that the antisymmetrized part of the Einstein-like field equations is identically zero, just as it happens in the theories with torsion and metricity \cite{FV1}. In fact, this was an expected result that restores equivalence between metric-affine and tetrad-affine formulations of the theory. 

As an application of the proposed theory, we have studied spinor fields in a Bianchi type-I (BI) spacetime arising from $f(\mathcal{Q})$ gravity. 
In cosmology, spinor fields have been mostly considered since the 1990s. They can drive the universe into accelerated expansion at early and late time, thus offering theoretical frameworks for both inflation and dark energy. Cosmologies sourced by fermions are even known to be able to avoid the initial cosmological
singularity \cite{Kovalyov,Obukhov,Saha1,Binetruy,Saha2,Picon,Saha3,Ribas1,Saha4,Boehmer1,Boehmer2,Ribas2,Rakhi,Rakhi2,VFC,Poplawski:2010jv,Ribas4,Ribas3,Poplawski:2011jz,Poplawski:2011wj,Magueijo:2012ug,Saha14,Souza,Devecchi,CVC,VCF,GSK}. Moreover, when dealing with spinor fields, anisotropic models of spacetime seem more appropriate (for example, think of the anisotropy induced by the spin four-vector). In connection with this, we have considered BI spacetimes as the simplest anisotropic spacetimes that generalize the spatially flat Friedmann-Lema\"{\i}tre-Robertson-Walker (FLRW) universe. Though FLRW models provide very accurate descriptions of the universe observed today, the latter could have undergone an anisotropic phase in its early epoch and might even become anisotropic in the future. Naturally, for consistency with the observed data, the initial anisotropic phase must undergo a subsequent isotropization process. This dynamical behavior can represent a selection rule for viable functions $f(\mathcal{Q})$. In this regard, we show that the functions of the kind $f(\mathcal{Q})=\alpha\/\mathcal{Q}^n$ can give rise to initially anisotropic universes that isotropize at late time and also undergo an accelerated expansion. 

The layout of the paper is the following. In Section \ref{sec.II}, we briefly introduce all the geometrical notions we will employ in the subsequent Sections. In Section \ref{sec.III}, we develop $f(\mathcal{Q})$ gravity in the presence of spinor fields, deducing and discussing the corresponding field equations. In Section \ref{sec.IV}, we focus on BI cosmological models, solving analytically the resulting field equations in the case $f(\mathcal{Q})=\alpha\/\mathcal{Q}^n$. We devote Section \ref{sec.V} to conclusions. Throughout the paper natural units ($\hbar=c=k_{B}=8\pi G=1$) and metric signature $(+,-,-,-)$ are used.

\section{Geometrical preliminaries}\label{sec.II}
Throughout the paper, spacetime (holonomic) indexes are indicated by Latin letters, while Lorentz (anholonomic) indexes by Greek letters; both sets of indexes run from $0$ to $3$. A metric tensor on the spacetime $M$ is denoted by $g_{ij}$ and an affine connection by $\Gamma_{ij}^{\;\;\;h}$. Any given affine connection $\Gamma_{ij}^{\;\;\;h}$ induces the corresponding covariant derivative, according to the notation
\begin{equation}\label{2.1}
\nabla_{\partial_i}\partial_j = \Gamma_{ij}^{\;\;\;h}\,\partial_h
\end{equation}
The torsion and the curvature tensors associated with a connection are defined as
\begin{equation}\label{2.2}
T_{ij}^{\;\;\;h}
=\Gamma_{ij}^{\;\;\;h}-\Gamma_{ji}^{\;\;\;h}
\end{equation}
and 
\be\label{2.2bis}
R^{h}_{\;\;kij}
=\partial_i\Gamma_{jk}^{\;\;\;h} - \partial_j\Gamma_{ik}^{\;\;\;h} +
\Gamma_{ip}^{\;\;\;h}\Gamma_{jk}^{\;\;\;p}-\Gamma_{jp}^{\;\;\;h}\Gamma_{ik}^{\;\;\;p}
\ee
The assignment of a metric tensor $g_{ij}$ allows us to introduce the non--metricity tensor of the connection $\Gamma_{ij}^{\;\;\;h}$, defined as
\be\label{2.3}
Q_{ijh}=\nab_ig_{jh}
\ee
The latter possesses two distinguished traces, denoted respectively by
\be\label{2.3bis}
Q_i=Q_{ji}^{\;\;\;j} \qquad {\rm and} \qquad q_i=Q_{ij}^{\;\;\;j}
\ee
By means of the torsion and non--metricity tensors \eqref{2.2} and \eqref{2.3}, any given connection may be espressed as
\begin{equation}\label{2.4}
\Gamma_{ij}^{\;\;\;h}=\tilde{\Gamma}_{ij}^{\;\;\;h}-K_{ij}^{\;\;\;h}-N_{ij}^{\;\;\;h}
\end{equation}
where $\tilde{\Gamma}_{ij}^{\;\;\;h}=\frac{1}{2}g^{hk}\left(\partial_ig_{jk}+\partial_jg_{ik}-\partial_kg_{ij}\right)$ is the Levi--Civita connection induced by the metric $g_{ij}$, while
\begin{equation}\label{2.5}
K_{ij}^{\;\;\;h}
=\frac{1}{2}\left(-T_{ij}^{\;\;\;h}+T_{j\;\;\;i}^{\;\;h}-T^{h}_{\;\;ij}\right)
\end{equation}
and
\be\label{2.6}
N_{ij}^{\;\;\;h}= \frac{1}{2}\left(Q_{ij}^{\;\;\;h}+Q_{ji}^{\;\;\;h}-Q^{h}_{\;\;ij}\right)
\ee
are the so called contorsion and the disformation tensors respectively.

In the next sections we shall consider $f(Q)$-gravity coupled with a spinor field; it is then useful describing the metric tensor $g_{ij}$ in terms of tetrad fields $e^{\mu}_{i}$: $g_{ij}=e^{\mu}_{i}e^{\mu}_{j}\eta_{\mu \nu}$, $\eta_{\mu\nu}$ denoting the Minkowskian metric with signature $(1,-1,-1,-1)$. Tetrad fields possess Lorentz indexes as well as spacetime indexes; they are defined by $e^{\mu}=e^{\mu}_{i}\,dx^{i}$ and, together with their dual fields $e_{\mu}=e_{\mu}^{i}\,\de /de{x^{i}}$, they satisfy the relations $e^{j}_{\mu}e^{\mu}_{i}=\delta^{j}_{i}$ and $e^{j}_{\mu}e^{\nu}_{j}=\delta^{\nu}_{\mu}$.

The adoption of a tetrad field may be seen as a change of local trivialization of the frame bundle over the spacetime $M$. As a consequence, the linear connection $\Gamma_{ij}^{\;\;\;h}$ gives rise to the corresponding spin connection
\be\label{2.7} 
\omega_{i\;\;\;\nu}^{\;\;\mu} = -e_\nu^k\partial_{i}e^{\mu}_{k} + e_{\nu}^{k}\Gamma_{ik}^{\;\;\;j}e^{\mu}_{j}
\ee
Relation \eqref{2.7} ensures that the full covariant derivative of the tetrad field $e^{\mu}_{i}$, with respect to both Latin and Greek indexes, vanishes; we have indeed
\begin{equation}\label{2.8}
\nabla_{j}e^{\mu}_{i} = \partial_{j}e^{\mu}_{i} + \omega_{i\;\;\;\nu}^{\;\;\mu}e^{\nu}_{i} - \Gamma_{ji}^{\;\;\;k}e^{\mu}_{k}=0
\end{equation}
As far as spin connection is concerned, the analogous decomposition of \eqref{2.4} assumes the form
\be\label{2.10}
\omega_{i\;\;\;\nu}^{\;\;\mu} = \tilde{\omega}_{i\;\;\;\nu}^{\;\;\mu} - K_{i\;\;\;\nu}^{\;\;\mu} - N_{i\;\;\;\nu}^{\;\;\mu}
\ee
where
\be\label{2.10bis}
\tilde{\omega}_{i\;\;\;\nu}^{\;\;\mu} = -e_\nu^k\partial_{i}e^{\mu}_{k} + e_{\nu}^{k}\tilde{\Gamma}_{ik}^{\;\;\;j}e^{\mu}_{j}
\ee
\be\la{2.11}
K_{i\;\;\;\nu}^{\;\;\mu} = K_{ij}^{\;\;\;h}e^\mu_h\/e^j_\nu
\ee
\be\la{2.12}
N_{i\;\;\;\nu}^{\;\;\mu} =N_{ij}^{\;\;\;h}e^\mu_h\/e^j_\nu
\ee
Following \cite{Poplawski}, given a spinor field $\psi$ we define its covariant derivative with respect to a general connection as
\be\la{2.13}
\nabla_i\psi =\partial_i\psi + \Omega_i\psi
\ee
with
\be\la{2.14}
\Omega_i= \frac{1}{8}\omega_i^{\;\;\mu\nu}\left[\gamma_\mu,\gamma_\nu\right]
\ee
In eq. \eqref{2.14}, Greek indexes are raised and lowered by the Minkowski metric $\eta_{\mu\nu}$, and $\gamma^\mu$ are a suitable choice of Dirac matrices satisfying the well-known relation $\{\gamma^\mu,\gamma^\nu\}=2\eta^{\mu\nu}$. 

As it is evident from Eq. \eqref{2.14}, the spinor covariant derivative involves antisymmetrized Lorentz indexes. About this, we recall that the antisymmetry of the Lorentz indexes amounts to the metric--compatibility of the affine connection, which, among other things, guarantees the vanishing of the covariant derivative of Dirac matrices; on the contrary, in the case of non--metricity, Dirac matrices have in general non zero covariant derivatives. To see this point in detail, we recall that the covariant derivative of Dirac matrices is expressed as
\be\la{2.15}
\nabla_i\gamma^\mu = \omega_{i\;\;\;\nu}^{\;\;\mu}\gamma^\nu + \left[\Omega_i,\gamma^\mu\right]
\ee
Making use of eq. \eqref{2.14} as well as of the algebraic identities
\be\la{2.16}
\left[\gamma^\mu,\left[\gamma^\sigma,\gamma^\tau\right]\right]=4\left(\gamma^\tau\/\eta^{\mu\sigma}-\gamma^\sigma\/\eta^{\mu\tau}\right)
\ee
from eq. \eqref{2.15} we get the the relation
\be\la{2.17}
\nabla_i\gamma^\mu = \omega_i^{\;\;(\mu\nu)}\gamma_\nu = -N_i^{\;\;(\nu\mu)}\gamma_\nu
\ee
with $\gamma_\nu=\gamma^\mu\eta_{\mu\nu}$. Denoting by $\gamma^i = \gamma^\mu\/e_\mu^i$ and taking the identity \eqref{2.8} into account, we have also
\be\la{2.18}
\nabla_i\gamma^j = - N_i^{\;\;(hj)}\gamma_h
\ee
where $\gamma_h =\gamma^k\/g_{kh}$.
\section{$f(\mathcal{Q})$-gravity coupled to a spinor field}\label{sec.III}
In this section, we shall consider $f(\mathcal{Q})$-gravity coupled with a spinor field $\psi$. In order to deal with spinor fields, we shall assume as gravitational fields a tetrad $e^\mu_i$ and an affine connection $\Gamma_{ij}^{\;\;\;h}$, defined on the spacetime $M$; the metric tensor on $M$ will be then expressed as $g_{ij}=e^\mu_ie^\nu_j\eta_{\mu\nu}$. 

We begin by introducing the non--metricity scalar
\be\label{3.1}
\mathcal{Q} = \frac{1}{4}Q_{hij}Q^{hij} - \frac{1}{2}Q_{hij}Q^{ijh} - \frac{1}{4} q_{h}q^{h} + \frac{1}{2}q_{h}Q^{h}
\ee
We notice that, after defining the quantity
\be\label{3.1bis}
P^{h}{}_{ij} = - \frac{1}{4} Q^{h}{}_{ij} + \frac{1}{4} Q_{ij}{}^{h} + \frac{1}{4} Q_{ji}{}^{h} + \frac{1}{4} q^{h} g_{ij} - \frac{1}{4} Q^{h} g_{ij} - \frac{1}{8} \delta^{h}_{i} q_{j} - \frac{1}{8}\delta^{h}_{j} q_{i}
\ee
the non--metricity scalar \eqref{3.1} can be expressed as
\be\label{3.1tris}
\mathcal{Q}=-Q_{hij}P^{hij}
\ee
The action functional of the theory is
\be\label{3.2}
\mathcal{A}(e^\mu_i,\Gamma_{ij}^{\;\;\;h},\psi)=\int \left[ \sqrt{-g} f(\mathcal{Q}) + \lambda_{h}{}^{kij}R^{h}{}_{kij} + \lambda_{h}{}^{ij}T_{ij}{}^{h} - \mathcal{L}_D\right]d^4x 
\ee
where $f(Q)$ is a given function of the scalar \eqref{3.1}, $\lambda_{h}{}^{kij}$ and $\lambda_{h}{}^{ij}$ are tensor densities playing the role of Lagrange multipliers, and $\mathcal{L}_D$ is the usual Dirac Lagrangian
\be\label{3.3}
\mathcal{L}_D =\sqrt{-g} \left[ \frac{i}{2} \left( \bar{\psi} \gamma^{i}\nabla_{i}\psi - \nabla_{i}\bar{\psi}\gamma^{i}\psi \right) - m\bar{\psi}\psi \right]
\ee
The field equations are derived by varying the action \eqref{3.2} with respect to tetrad, affine connection, spinor field, and Lagrange multipliers. More specifically, variations with respect to the Lagrange multipliers give rise to the two constraints
\be\label{3.4}
R^{h}{}_{kij}=0 \qquad {\rm and} \qquad T_{ij}{}^{h}=0
\ee
In order to carry out variations with respect to the spinor field, it is convenient making use of the identity \eqref{2.10} so to express the covariant spinor derivative in the form
\be\label{3.4.1}
\nabla_i\psi= \tilde{\nabla}_i\psi + \frac{1}{8}N_i^{\;\;\nu\mu}\left[\gamma_\mu\gamma_\nu\right]\psi \qquad {\rm and} \qquad 
\nabla_i\bar\psi = \tilde{\nabla}_i\bar\psi - \frac{1}{8}\bar\psi\/N_i^{\;\;\nu\mu}\left[\gamma_\mu\gamma_\nu\right]
\ee
where the constraint $T_{ij}^{\;\;\;h}=0$ has been taken into account. After inserting eqs. \eqref{3.4.1} into the Dirac Lagrangian \eqref{3.3}, integrating by parts and discharging a suitable boundary term, straightforward calculations together with the identity \eqref{2.16} allow us to get the Dirac equations
\be\label{3.4.2}
i\gamma^h\/\nabla_h\psi + \frac{i}{4}q_h\gamma^h\psi - \frac{i}{4}Q_h\gamma^h\psi - m\psi =0 \quad {\rm and} \quad
i\/\nabla_h\bar\psi\gamma^h + \frac{i}{4}q_h\bar\psi\gamma^h - \frac{i}{4}Q_h\bar\psi\gamma^h + m\bar\psi =0
\ee
By varying with respect to the connection (see \cite{Harko1} for details), we obtain the equations
\be\label{3.5}
2 \nabla_{p} \lambda_{h}{}^{jip} + 2 \lambda_{h}{}^{ij} + 4\sqrt{-g} f' P^{ij}{}_{h} = \Phi^{ij}{}_{h}
\ee
where $f'=\frac{\partial f}{\partial \mathcal{Q}}$ and $\Phi^{ij}{}_{h}= \frac{\delta\mathcal{L}_D}{\delta\Gamma_{ij}{}^{h}}$. As we will see later, it is not necessary to elaborate further on this last equation.

In order to perform the variation with respect to the tetrad field, we exploit the results already present in the literature \cite{Harko1, Koivisto1} together with the identities
\be\label{3.6}
\frac{\partial e}{\partial e^\mu_i}= e\/e^i_\mu \qquad {\rm and} \qquad \frac{\partial e^j_\nu}{\partial e^\mu_i}=-e^i_\nu\/e^j_\mu
\ee
where $e=\sqrt{-g}$. On one hand from \cite{Harko1, Koivisto1} we have
\be\label{3.7}
\frac{\delta \sqrt{-g}f(\mathcal{Q})}{\delta\/g^{ij}}\delta\/g^{ij}= \sqrt{-g}\left[\frac{2}{\sqrt{-g}}\nabla_{k} \left( \sqrt{-g} f' P^{k}{}_{ij} \right) + \frac{1}{2}g_{ij}f(\mathcal{Q}) + f' \left( P_{i|ab}Q_{|j}{}^{ab} - 2 Q^{ab}{}_{(i|}P_{ab|j} \right)\right]\delta\/g^{ij}
\ee
On the other hand, making use of eqs. \eqref{3.6}, it is easily seen that
\be\label{3.8}
\frac{\delta\/g^{ij}}{\delta e^\tau_h}= -2g^{jh}e^i_\tau
\ee
Collecting the above results, the variation of the gravitational Lagrangian with respect to the tetrad field may be expressed as  
\be\label{3.9}
\frac{\delta \sqrt{-g}f(\mathcal{Q})}{\delta\/e^\tau_h} = -2\sqrt{-g}\left[\frac{2}{\sqrt{-g}}\nabla_{k} \left( \sqrt{-g} f' P^{k}{}_{ij} \right) + \frac{1}{2}g_{ij}f(\mathcal{Q}) + f' \left( P_{iab}Q_{j}{}^{ab} - 2 Q^{ab}{}_{i}P_{abj} \right)\right]\/g^{jh}e^i_\tau
\ee
For later use, saturating eq. \eqref{3.9} by $e^\tau_q$ and lowering the index $h$, we obtain the equivalent identity
\be\la{3.9bis}
\frac{\delta \sqrt{-g}f(\mathcal{Q})}{\delta\/e^\tau_h}e^\tau_q\/g_{hs} = -2\sqrt{-g}\left[\frac{2}{\sqrt{-g}}\nabla_{k} \left( \sqrt{-g} f' P^{k}{}_{qs} \right) + \frac{1}{2}g_{qs}f(\mathcal{Q}) + f'\left(P_{qab}Q_{s}{}^{ab} - 2 Q^{ab}{}_{q}P_{abs} \right)\right]
\ee
The variation of Dirac Lagrangian with respect to $e^\tau_h$ deserves a little more attention. First of all, we observe that eq. \eqref{2.7} allows us to represent the full covariant derivative of the variation $\delta e^\tau_h$ in the form
\be\label{3.10}
\nabla_i\left(\delta e^\tau_h\right) = \partial_i\left(\delta e^\tau_h\right) - \partial_i\left(e^\tau_p\right)e^p_\gamma\delta\/e^\gamma_h - \Gamma_{ih}^{\;\;\;s}\delta e^\tau_s + \Gamma_{ip}^{\;\;\;q}e^\tau_q\/e^p_\gamma\delta\/e^\gamma_h 
\ee
With the identity \eqref{3.10} in mind, after performing the variation of the Dirac Lagrangian with respect to the tetrad field, we get the expression
\be\label{3.11}
\frac{\delta \mathcal{L}_D}{\delta e^\tau_h}\delta e^\tau_h =e^h_\tau\mathcal{L}_D\delta e^\tau_h -
\frac{i}{2}e\left[\bar\psi\gamma^\mu\/e_\mu^h\/e^i_\tau\/\nabla_i\psi - \nabla_i\bar\psi\gamma^\mu\/e_\mu^h\/e^i_\tau\psi\right]\delta e^\tau_h - 
\frac{i}{16}e\nabla_i\left(\delta e^\tau_h\right)\bar\psi\left\{\gamma^i,\left[\gamma_\tau,\gamma^h\right]\right\}\psi
\ee
where, on the right hand side, the presence of the last addendum is due to the fact that we are using the tetrad field and the affine connection to represent the spin connection \eqref{2.7}, involved in the spinor covariant derivative \eqref{2.13}. Now, taking eqs. \eqref{2.4}, \eqref{2.10}, \eqref{2.10bis} and \eqref{2.12} as well as the constraint $T_{ij}^{\;\;\;h}=0$ into account, we have the identity
\be\label{3.12}
\nabla_i\left(\delta\/e^\tau_h\right)= \tilde{\nabla}_i\left(\delta\/e^\tau_h\right) + N_{ih}^{\;\;\;s}\delta\/e^\tau_s - N_{i\;\;\;\sigma}^{\;\;\tau}\delta\/e^\sigma_h
\ee
where $\tilde{\nabla}_i$ indicates full covariant derivative with respect to Levi--Civita affine and spin connections. In view of eq. \eqref{3.12}, after some calculations we end up with the further identity
\be\label{3.13}
\begin{split}
- \frac{i}{16}e\nabla_i\left(\delta e^\tau_h\right)\bar\psi\left\{\gamma^i,\left[\gamma_\tau,\gamma^h\right]\right\}\psi = - \frac{i}{16}e\tilde{\nabla}_i\left(\delta e^\tau_h\bar\psi\left\{\gamma^i,\left[\gamma_\tau,\gamma^h\right]\right\}\psi\right) + \frac{i}{16}e\tilde{\nabla}_i\left(\bar\psi\left\{\gamma^i,\left[\gamma_\tau,\gamma^h\right]\right\}\psi\right)\delta e^\tau_h \\
- \frac{i}{16}eN_{is}^{\;\;\;h}\bar\psi\left\{\gamma^i,\left[\gamma_\tau,\gamma^s\right]\right\}\psi\delta e^\tau_h + \frac{i}{16}eN_{i\,\,\,\tau}^{\;\;\sigma}\bar\psi\left\{\gamma^i,\left[\gamma_\sigma,\gamma^h\right]\right\}\psi\delta e^\tau_h
\end{split}
\ee
Notice that the first addendum on the right-hand side of eq. \eqref{3.13} is a divergence that leads to a boundary term.
Moreover, still using eqs. \eqref{2.4}, \eqref{2.10}, \eqref{2.10bis} and \eqref{2.12}, it is an easy matter to verify the relation
\be\label{3.14}
\begin{split}
\frac{i}{16}e\tilde{\nabla}_i\left(\bar\psi\left\{\gamma^i,\left[\gamma_\tau,\gamma^h\right]\right\}\psi\right)
- \frac{i}{16}eN_{is}^{\;\;\;h}\bar\psi\left\{\gamma^i,\left[\gamma_\tau,\gamma^s\right]\right\}\psi + \frac{i}{16}eN_{i\,\,\,\tau}^{\;\;\sigma}\bar\psi\left\{\gamma^i,\left[\gamma_\sigma,\gamma^h\right]\right\}\psi=\\
\frac{i}{16}e\nabla_i\left(\bar\psi\left\{\gamma^i,\left[\gamma_\tau,\gamma^h\right]\right\}\psi\right) + \frac{i}{16}eN_{is}^{\;\;\;i}\bar\psi\left\{\gamma^s,\left[\gamma_\tau,\gamma^h\right]\right\}\psi
\end{split}
\ee
Replacing the content of eqs. \eqref{3.13} and \eqref{3.14} into eq. \eqref{3.11}, we obtain the identity
\be\la{3.15}
\frac{\delta \mathcal{L}_D}{\delta e^\tau_h}=e^h_\tau\/\mathcal{L}_D -
\frac{i}{2}e\left[\bar\psi\gamma^\mu\/e_\mu^h\/e^i_\tau\/\nabla_i\psi - \nabla_i\bar\psi\gamma^\mu\/e_\mu^h\/e^i_\tau\psi\right] +
\frac{i}{16}e\nabla_i\left(\bar\psi\left\{\gamma^i,\left[\gamma_\tau,\gamma^h\right]\right\}\psi\right) + \frac{i}{16}eN_{is}^{\;\;\;i}\bar\psi\left\{\gamma^s,\left[\gamma_\tau,\gamma^h\right]\right\}\psi
\ee
which, saturated by $e^\tau_q$, yields
\be\la{3.16}
\frac{\delta \mathcal{L}_D}{\delta e^\tau_h}e^\tau_q = \delta^h_q\/\mathcal{L}_D -
\frac{i}{2}e\left[\bar\psi\gamma^h\/\nabla_q\psi - \nabla_q\bar\psi\gamma^h\psi\right] +
\frac{i}{16}e\nabla_i\left(\bar\psi\left\{\gamma^i,\left[\gamma_q,\gamma^h\right]\right\}\psi\right) + \frac{i}{16}eN_{is}^{\;\;\;i}\bar\psi\left\{\gamma^s,\left[\gamma_q,\gamma^h\right]\right\}\psi
\ee 
Expression \eqref{3.16} can be further elaborated and simplified, making use of the Dirac equations \eqref{3.4.2}. First, it is an easy matter to verify that the Dirac Lagrangian $\mathcal{L}_D$ vanishes on shell. Thus, the first addendum on the right hand side of eq. \eqref{3.16} may be omitted, as usual. Moreover, once the index $h$ is lowered, the antisymmetric part of whole right hand side of eq. \eqref{3.16} vanishes as well. To see this point, we analyze in detail the divergence term
\be\la{3.17}
\begin{split}
\frac{i}{16}e\nabla_i\left(\bar\psi\left\{\gamma^i,\left[\gamma_q,\gamma^h\right]\right\}\psi\right) = 
\frac{i}{16}e\left(\nabla_i\bar\psi\right)\gamma^i\left[\gamma_q,\gamma^h\right]\psi + 
\frac{i}{16}e\bar\psi\gamma^i\left[\gamma_q,\gamma^h\right]\left(\nabla_i\psi\right) +
\frac{i}{16}e\left(\nabla_i\bar\psi\right)\left[\gamma_q,\gamma^h\right]\gamma^i\psi +\\
\frac{i}{16}e\bar\psi\left[\gamma_q,\gamma^h\right]\gamma^i\left(\nabla_i\psi\right) +
\frac{i}{16}e\bar\psi\left(\nabla_i\left\{\gamma^i,\left[\gamma_q,\gamma^h\right]\right\}\right)\psi
\end{split}
\ee
By adding and subtracting the terms $\frac{i}{16}e\bar\psi\left[\gamma_q,\gamma^h\right]\gamma^i\left(\nabla_i\psi\right)$ and $\frac{i}{16}e\left(\nabla_i\bar\psi\right)\gamma^i\left[\gamma_q,\gamma^h\right]\psi$ in eq. \eqref{3.17}, we obtain the following expression
\be\la{3.18}
\begin{split}
\frac{i}{16}e\nabla_i\left(\bar\psi\left\{\gamma^i,\left[\gamma_q,\gamma^h\right]\right\}\psi\right) = 
\frac{i}{2}e\left(- \frac{1}{8}\left(\nabla_i\bar\psi\right)\left[\gamma^i,\left[\gamma_q,\gamma^h\right]\right]\psi + 
\frac{1}{8}\bar\psi\left[\gamma^i,\left[\gamma_q,\gamma^h\right]\right]\left(\nabla_i\psi\right)\right) \\
+ \frac{1}{8}e\left(\left(i\nabla_i\bar\psi\gamma^i\right)\left[\gamma_q,\gamma^h\right]\psi + 
\bar\psi\left[\gamma_q,\gamma^h\right]\left(i\gamma^i\nabla_i\psi\right)\right) +
\frac{i}{16}e\bar\psi\left(\nabla_i\left\{\gamma^i,\left[\gamma_q,\gamma^h\right]\right\}\right)\psi
\end{split}
\ee
By employing the Dirac equations \eqref{3.4.2} and the identity \eqref{2.16}, from eq. \eqref{3.18} we easily get
\be\la{3.19}
\begin{split}
\frac{i}{16}e\nabla_i\left(\bar\psi\left\{\gamma^i,\left[\gamma_q,\gamma^h\right]\right\}\psi\right) =
\frac{i}{4}e\left(\bar\psi\gamma^h\nabla_q\psi - \bar\psi\gamma_q\nabla^h\psi - \nabla_q\bar\psi\gamma^h\psi + \nabla^h\bar\psi\gamma_q\psi\right) \\
+ \frac{i}{32}e\left(Q_{ji}^{\;\;\;j} - Q_{ij}^{\;\;\;j}\right)\bar\psi\left\{\gamma^i,\left[\gamma_q,\gamma^h\right]\right\}\psi +
\frac{i}{16}e\bar\psi\left(\nabla_i\left\{\gamma^i,\left[\gamma_q,\gamma^h\right]\right\}\right)\psi
\end{split}
\ee
Except for inessential multiplying factors, eq. \eqref{3.19} represents the conservation law of the spin density in the current theory. Compared to that holding in Einstein-Cartan-like theories (for example, see \cite{FV1}), there are evident differences due to the explicit presence of the non-metricity tensor and to the fact that the covariant derivatives of Dirac matrices are no longer zero. 

To proceed further, we must to elaborate the term $\frac{i}{16}e\bar\psi\left(\nabla_i\left\{\gamma^i,\left[\gamma_q,\gamma^h\right]\right\}\right)\psi$. To this end, making use of relation \eqref{2.18}, we have 
\be\label{3.20}
\begin{split}
\frac{i}{16}e\bar\psi\left(\nabla_i\left\{\gamma^i,\left[\gamma_q,\gamma^h\right]\right\}\right)\psi =
- \frac{i}{32}eN_i^{\;\;pi}\bar\psi\left\{\gamma_p,\left[\gamma_q,\gamma^h\right]\right\}\psi
- \frac{i}{32}eN_i^{\;\;ip}\bar\psi\left\{\gamma_p,\left[\gamma_q,\gamma^h\right]\right\}\psi\\
- \frac{i}{32}eQ_{ipq}\bar\psi\left\{\gamma^i,\left[\gamma^h,\gamma^p\right]\right\}\psi
- \frac{i}{32}eQ_{ip}^{\;\;\;h}\bar\psi\left\{\gamma^i,\left[\gamma_q,\gamma^p\right]\right\}\psi
\end{split}
\ee
At this point, in view of eqs. \eqref{3.19} and \eqref{3.20} as well as of the identities $N_{is}^{\;\;\;i}=\frac{1}{2}Q_{si}^{\;\;\;i}$ and 
$N_{i\;\;p}^{\;\;i}=Q_{i\;\;p}^{\;\;i} - \frac{1}{2}Q_{pi}^{\;\;\;i}$, we are able to express the sum of the last two terms on the right hand side of eq. \eqref{3.16} in the form
\be\la{3.21}
\begin{split}
\frac{i}{16}e\nabla_i\left(\bar\psi\left\{\gamma^i,\left[\gamma_q,\gamma^h\right]\right\}\psi\right) + \frac{i}{16}eN_{is}^{\;\;\;i}\bar\psi\left\{\gamma^s,\left[\gamma_q,\gamma^h\right]\right\}\psi = 
\frac{i}{4}e\left(\bar\psi\gamma^h\nabla_q\psi - \bar\psi\gamma_q\nabla^h\psi - \nabla_q\bar\psi\gamma^h\psi + \nabla^h\bar\psi\gamma_q\psi\right)\\
- \frac{i}{32}eQ_{ipq}\bar\psi\left\{\gamma^i,\left[\gamma^h,\gamma^p\right]\right\}\psi
- \frac{i}{32}eQ_{ip}^{\;\;\;h}\bar\psi\left\{\gamma^i,\left[\gamma_q,\gamma^p\right]\right\}\psi
\end{split}
\ee
Inserting eq. \eqref{3.21} into eq. \eqref{3.16} and lowering the index $h$, we obtain the final expression 
\be\la{3.22}
\frac{\delta \mathcal{L}_D}{\delta e^\tau_h}e^\tau_q\/g_{hs}= - \frac{i}{2}e\left(\bar\psi\gamma_{(s}\nabla_{q)}\psi - \nabla_{(q}\bar\psi\gamma_{s)}\psi\right) - \frac{i}{32}eQ_{ipq}\bar\psi\left\{\gamma^i,\left[\gamma_s,\gamma^p\right]\right\}\psi
- \frac{i}{32}eQ_{ips}\bar\psi\left\{\gamma^i,\left[\gamma_q,\gamma^p\right]\right\}\psi
\ee
To conclude, by equating eq. \eqref{3.9bis} with eq. \eqref{3.22} and dividing by $\sqrt{-g}=e$, we get the explicit form of the field equations deduced by varying with respect to the tetrad field, namely
\be\la{3.23}
\begin{split}
\frac{2}{\sqrt{-g}}\nabla_{k} \left( \sqrt{-g} f' P^{k}{}_{qs} \right) + \frac{1}{2}g_{qs}f(\mathcal{Q}) + f'\left(P_{qab}Q_{s}{}^{ab} - 2 Q^{ab}{}_{q}P_{abs} \right) = \\
\frac{i}{4}\left(\bar\psi\gamma_{(s}\nabla_{q)}\psi - \nabla_{(q}\bar\psi\gamma_{s)}\psi\right) + 
\frac{i}{64}Q_{ipq}\bar\psi\left\{\gamma^i,\left[\gamma_s,\gamma^p\right]\right\}\psi
+ \frac{i}{64}Q_{ips}\bar\psi\left\{\gamma^i,\left[\gamma_q,\gamma^p\right]\right\}\psi
\end{split}
\ee
where both sides of the equation \eqref{3.23} are symmetric in the indexes $q$ and $s$. For the sake of completeness, we also write the fully developed expression of eq. \eqref{3.5}
\be\label{3.24}
2 \nabla_{p} \lambda_{h}{}^{jip} + 2 \lambda_{h}{}^{ij} + 4\sqrt{-g} f' P^{ij}{}_{h} = -\frac{i\sqrt{-g}}{16}\bar\psi\left\{\gamma^i,\left[\gamma^j,\gamma_h\right]\right\}\psi
\ee
Eqs. \eqref{3.23} and \eqref{3.24}, together with eqs. \eqref{3.4} and \eqref{3.4.2}, represent the totality of the field equations. As far as the search for solutions is concerned, it is worth noticing that the constraints \eqref{3.4} imply the existence of local coordinates, the so-called coincidence gauge, in which the connection coefficients $\Gamma_{ij}^{\;\;\;h}$ vanish. Therefore, in the coincidence gauge eqs. \eqref{3.24} are only intended for the determination of the Lagrange multipliers. However, this step is not strictly necessary because Lagrange multipliers do not enter eqs. \eqref{3.4.2} and \eqref{3.23} and so their determination becomes irrelevant. Thus, adopting the coincidence gauge, eqs. \eqref{3.24} can be neglected while eqs. \eqref{3.4} are trivially satisfied; this result allows us to simplify significantly the search for solutions, as now we need to solve only eqs. \eqref{3.4.2} and \eqref{3.23}. Of course, this is no longer true if one chooses to work with other gauges.   

\section{Bianchi--I cosmological models}\label{sec.IV}
In the coincidence gauge $\Gamma_{ij}^{\;\;\;h}=0$, we assume a Bianchi type I metric of the form
\begin{equation}\la{4.1}
ds^2 = dt^2 - a^2(t)dx^2 - b^2(t)dy^2 - c^2(t)dz^2
\end{equation}
describing a homogeneous and anisotropic universe. The components of the tetrad field associated with the line element \eqref{4.1} are expressed as
\begin{subequations}\label{4.2}
\begin{equation}\la{4.2a}
e^\mu_0=\delta^\mu_0, \quad e^\mu_1 = a(t)\delta^\mu_1, \quad e^\mu_2 = b(t)\delta^\mu_2, \quad e^\mu_3 = c(t)\delta^\mu_3 \qquad \mu =0,1,2,3
\end{equation}
with inverse relations given by
\begin{equation}\la{4.2b}
e^0_\mu = \delta^0_\mu, \quad e^1_\mu = \frac{1}{a(t)}\delta^1_\mu, \quad e^2_\mu = \frac{1}{b(t)}\delta^2_\mu, \quad e^3_\mu = \frac{1}{c(t)}\delta^3_\mu \qquad \mu =0,1,2,3
\end{equation}
\end{subequations}
Under the conditions $\Gamma_{ij}^{\;\;\;h}=0$, the only non-zero components of the non-metricity tensor \eqref{2.3} are seen to be
\begin{equation}\la{4.4}
Q_{011}=-2a\dot{a}, \quad Q_{022}=-2b\dot{b}, \quad Q_{033}=-2c\dot{c}
\end{equation}
and the traces \eqref{2.3bis} are respectively expressed as 
\be\la{4.4bis}
Q_i=0 \quad {\rm and} \quad q_i=2\frac{\dot\tau}{\tau}\delta^0_i \qquad i=0,\ldots,3
\ee
where we have defined the spatial volume of the universe as $\tau:=abc$. Analogously, the non--zero components of the tensor $P^{h}{}_{ij}$ \eqref{3.1bis} result to be
\begin{subequations}\label{4.5}
\begin{equation}\la{4.5a}
P^{0}{}_{11}=P_{011}=\frac{1}{2}a\dot{a} - \frac{1}{2} \frac{\dot{\tau}}{\tau}a^2 \quad
P^{0}{}_{22}=P_{022}=\frac{1}{2}b\dot{b} - \frac{1}{2} \frac{\dot{\tau}}{\tau}b^2 \quad
P^{0}{}_{33}=P_{033}=\frac{1}{2}c\dot{c} - \frac{1}{2} \frac{\dot{\tau}}{\tau}c^2
\end{equation}
\begin{equation}\label{4.5b}
P^{1}{}_{10}=P^{1}{}_{01}=- \frac{1}{4}\frac{\dot{\tau}}{\tau} + \frac{1}{2}\frac{\dot{a}}{a} \quad
P^{2}{}_{20}=P^{2}{}_{02}=- \frac{1}{4}\frac{\dot{\tau}}{\tau} + \frac{1}{2}\frac{\dot{b}}{b} \quad
P^{3}{}_{30}=P^{3}{}_{03}=- \frac{1}{4}\frac{\dot{\tau}}{\tau} + \frac{1}{2}\frac{\dot{c}}{c} 
\end{equation}
\end{subequations}
In view of eqs. \eqref{4.4} and \eqref{4.5}, the non--metricity scalar assumes the form
\begin{equation}\la{4.5bis}
\mathcal{Q} = - P^{0ii}Q_{0ii} = \left( \frac{\dot{a}^2}{a^2} + \frac{\dot{b}^2}{b^2} + \frac{\dot{c}^2}{c^2} \right) - \left( \frac{\dot{\tau}}{\tau} \right)^2 = -2 \left( \frac{\dot{a}\dot{b}}{ab} + \frac{\dot{a}\dot{c}}{ac} +\frac{\dot{b}\dot{c}}{bc} \right) 
\end{equation}
Moreover, homogeneity and coincidence gauge assumptions together with eqs. \eqref{2.7}, \eqref{2.13}, \eqref{2.14} and \eqref{4.2} yield the identities
\be\la{4.6}
\nabla_0\psi=\dot\psi, \quad \nabla_A\psi=0 \qquad A=1,\ldots,3
\ee
After inserting the content of eqs. \eqref{4.4bis} and \eqref{4.6} into eqs. \eqref{3.4.2}, Dirac equations assume the form
\begin{subequations}\la{4.7}
\begin{equation}\la{4.7a}
i\gamma^{0}\partial_{0}\psi + \frac{i}{2}\frac{\dot{\tau}}{\tau}\gamma^{0}\psi - m\psi = 0
\end{equation}
\begin{equation}\la{4.7b}
i\partial_{0}\bar{\psi}\gamma^{0} + \frac{i}{2}\frac{\dot{\tau}}{\tau}\bar{\psi}\gamma^{0} + m\bar{\psi} = 0.
\end{equation}
\end{subequations}
Eqs. \eqref{4.7} can be easily integrated; adopting the Dirac representation for the matrices $\gamma^\mu$, they possess solutions of the form
\begin{eqnarray}\la{4.8}
&\psi=\frac{1}{\sqrt{\tau}}\left(\begin{tabular}{c}
$c_{1}e^{-imt}$\\
$c_{2}e^{-imt}$\\
$c_{3}e^{imt}$\\
$c_{4}e^{imt}$
\end{tabular}\right)
\end{eqnarray}
where $c_i$, $i=1,\ldots,4$, are suitable integration constants. Moreover, from Dirac equations \eqref{4.7} or also from their solutions \eqref{4.8}, it is easily seen the following relation necessarily holds 
\begin{equation}\la{4.9}
\frac{d}{dt}\left( \tau\bar{\psi}\psi \right) = 0 \quad \Longleftrightarrow \quad \bar{\psi}\psi = \frac{K}{\tau}
\end{equation}
with $K = |c_{1}|^{2} + |c_{2}|^{2} - |c_{3}|^{2} - |c_{4}|^{2}$.

As for eqs. \eqref{3.23}, due to eqs. \eqref{4.6} and \eqref{4.7}, the only non zero component of the tensor $\frac{i}{4}\left(\bar\psi\gamma_{(s}\nabla_{q)}\psi - \nabla_{(q}\bar\psi\gamma_{s)}\psi\right)$ is given by 
\begin{equation}\la{4.10}
\Sigma_{00} = \frac{i}{4} \left( \bar{\psi}\gamma_{0}\nabla_{0}\psi - \nabla_{0}\bar{\psi}\gamma_{0}\psi \right) = \frac{1}{2} m \bar{\psi}\psi.
\end{equation}
In view of this and making use of eqs. \eqref{4.4} and \eqref{4.5}, a direct calculation shows that field equations \eqref{3.23} assume the explicit form
\begin{subequations}\la{4.11}
\begin{equation}\la{4.11a} 
\frac{1}{2} f + 2f' \left( \frac{\dot{a}\dot{b}}{ab} + \frac{\dot{b}\dot{c}}{bc} + \frac{\dot{a}\dot{c}}{ac} \right) = \frac{1}{2}m\bar{\psi}\psi ,
\end{equation}
\begin{equation}\la{4.11b}
\dot{f}' \left( - \frac{\dot{a}}{a} + \frac{\dot{\tau}}{\tau} \right) + f' \left( \frac{\ddot{b}}{b} + \frac{\ddot{c}}{c} + \frac{\dot{a}\dot{b}}{ab} + \frac{\dot{a}\dot{c}}{ac} + 2 \frac{\dot{b}\dot{c}}{bc} \right) + \frac{1}{2}f =0
\end{equation}
\begin{equation}\la{4.11c}
\dot{f}' \left( - \frac{\dot{b}}{b} + \frac{\dot{\tau}}{\tau} \right) + f' \left( \frac{\ddot{a}}{a} + \frac{\ddot{c}}{c} + \frac{\dot{a}\dot{b}}{ab} + \frac{\dot{b}\dot{c}}{bc} + 2 \frac{\dot{a}\dot{c}}{ac} \right) + \frac{1}{2}f =0
\end{equation}
\begin{equation}\la{4.11d}
\dot{f}' \left( - \frac{\dot{c}}{c} + \frac{\dot{\tau}}{\tau} \right) + f' \left( \frac{\ddot{a}}{a} + \frac{\ddot{b}}{b} + \frac{\dot{a}\dot{c}}{ac} + \frac{\dot{b}\dot{c}}{bc} + 2 \frac{\dot{a}\dot{b}}{ab} \right) + \frac{1}{2}f =0
\end{equation}
\end{subequations}
\begin{subequations}\la{4.12}
\be\la{4.12a}
\left(\dot{a}b - a\dot{b}\right)\bar\psi\gamma^5\gamma_3\psi =0
\ee
\be\la{4.12b}
\left(\dot{a}c - a\dot{c}\right)\bar\psi\gamma^5\gamma_2\psi =0
\ee
\be\la{4.12c}
\left(\dot{b}c - b\dot{c}\right)\bar\psi\gamma^5\gamma_1\psi =0
\ee
\end{subequations}
where eqs. \eqref{4.11} and \eqref{4.12} derive from the diagonal and the off--diagonal part of eqs. \eqref{3.23} respectively. It is interesting to note that the conditions \eqref{4.12} are identical to those that arise also in $f(R)$ theories with torsion \cite{VFC}. In that case, however, these conditions stem directly from the Dirac tensor, while in the present case they originate from the additional terms appearing in the second member of eq. \eqref{3.23}.

Eqs. \eqref{4.12} are automatically satisfied in the case of an isotropic universe. Instead  anisotropic spacetimes imply stringent constraints on the spinor field i.e. 
\begin{equation}
\bar\psi\gamma^5\gamma^1\psi= \bar\psi\gamma^5\gamma^2\psi=\bar\psi\gamma^5\gamma^3\psi =0
\end{equation} 
In this case, the orthogonality between the current four-vector $\bar\psi\gamma^\mu\psi$ and the spin four-vector $\bar\psi\gamma^5\gamma^\mu\psi$ implies that the time component $\bar\psi\gamma^5\gamma^0\psi$ has to be zero. In fact, if $\bar\psi\gamma^0\psi$ were allowed to vanish, then the whole spinor field would be zero, but the vanishing of the entire spin four-vector implies the condition $\bar\psi\psi=0$. It follows that, in the anisotropic case, the spinor field does not enter the gravitational equations \eqref{4.11} which become identical to the ones we would have in vacuo.  
Of course, there are also intermediate situations where, in the face of a partial isotropy, some constraints persist on the spin four-vector: for instance, the case $a=b$ and $\bar\psi\gamma^5\gamma^1\psi=\bar\psi\gamma^5\gamma^2\psi=0$. Anyway, it is clear that any constraints imposed on the four-vector $\bar\psi\gamma^5\gamma^\mu\psi$ translate into restrictions on the admissible values of the integration constants $c_i$ appearing in eq. \eqref{4.8}. For example, the sets of constants $(c_1=e^{i\theta}c_4, c_2=c_3=0)$ or $(c_2=e^{i\theta}c_3, c_1=c_4=0)$ make the requirement $\bar\psi\gamma^5\gamma^\mu\psi=0$ satisfied; the less restrictive choice $(c_2=c_3=0)$ ensure the weaker condition $\bar\psi\gamma^5\gamma^1\psi=\bar\psi\gamma^5\gamma^2\psi=0$.

In order to discuss and solve eqs. \eqref{4.11}, we can follow the lines traced in previous works (see for example \cite{Saha2,VFC}) and, after expressing every dynamical variable as a function of the spatial volume $\tau$, try to get a solving equation for the unknown $\tau$ itself. To this end, we preliminarily observe that eq. \eqref{4.11a} can be rewritten as
\be\la{4.13}
\frac{1}{2}f(\mathcal{Q}) - f'(\mathcal{Q})\mathcal{Q} = \frac{mK}{2\tau}
\ee 
where eqs. \eqref{4.5bis} and \eqref{4.9} have been employed. Eq. \eqref{4.13} highlights how the contribution of the spinor field actually reduces to that of a cosmological dust.

Now, given the function $f(\mathcal{Q})$ and with the exception of some pathological cases, in general from eq. \eqref{4.13} we may derive the expression of the non--metricity scalar in terms of $\tau$, i.e. $\mathcal{Q}=\mathcal{Q}\left(\tau\right)$. In view of this, by subtracting eq. \eqref{4.11b} from eq. \eqref{4.11c} and from eq. \eqref{4.11d} separately, we obtain the two equations
\begin{subequations}\la{4.14}
\begin{equation}\la{4.14a}
\frac{d}{dt} \left[ f' \tau \left( \frac{\dot{a}}{a} - \frac{\dot{b}}{b} \right) \right] =0 
\end{equation}
\begin{equation}\la{4.14b}
\frac{d}{dt} \left[ f' \tau \left( \frac{\dot{a}}{a} - \frac{\dot{c}}{c} \right) \right] =0 
\end{equation}
\end{subequations} 
which in turn imply the relations
\begin{subequations}\la{4.15}
\begin{equation}\la{4.15a}
\frac{a}{b} = e^{d_{2}}\exp \int \frac{d_{1}}{f' \tau} dt
\end{equation}
\begin{equation}\la{4.15b}
\frac{a}{c} = e^{g_{2}} \exp \int \frac{g_{1}}{f' \tau} dt
\end{equation}
\end{subequations}
with $d_1$, $d_2$, $g_1$ and $g_2$ suitable integration constants and where $f'(\mathcal{Q}(\tau))$ is function of $\tau$.

If we could now get an equation for the variable $\tau$ alone, the systems of field equations would be entirely worked out. This goal may be achieved through a suitable combination of eqs. \eqref{4.11}. Indeed, by summing eqs. \eqref{4.11b}, \eqref{4.11c} and \eqref{4.11d} each to other and subtracting eq. \eqref{4.11a} multiplied by $3$, we get the final dynamical equation for the unknown $\tau$  
\begin{equation}\la{4.16}
\frac{2}{\tau}\frac{d}{dt}\left(f'\dot{\tau}\right) + 3f' \mathcal{Q} = - \frac{3mK}{2\tau}  
\end{equation}
Once eq. \eqref{4.16} was solved, from eqs. \eqref{4.15} together with the relation $\tau=abc$, we would have the expressions for the scale factors
\begin{subequations}\la{4.17}
\be\la{4.17a}
a(t)= \tau^{\frac{1}{3}}e^{\frac{\left(d_2+g_2\right)}{3}}e^{\frac{\left(d_1+g_1\right)}{3}\int^t_{t_0}{\frac{dt}{f'\tau}}}
\ee
\be\la{4.17b}
b(t)= \tau^{\frac{1}{3}}e^{\frac{\left(-2d_2+g_2\right)}{3}}e^{\frac{\left(-2d_1+g_1\right)}{3}\int^t_{t_0}{\frac{dt}{f'\tau}}}
\ee
\be\la{4.17c}
c(t)= \tau^{\frac{1}{3}}e^{\frac{\left(d_2-2g_2\right)}{3}}e^{\frac{\left(d_1-2g_1\right)}{3}\int^t_{t_0}{\frac{dt}{f'\tau}}}
\ee
\end{subequations} 
With the only exception $f\left(\mathcal{Q}\right)=\alpha\sqrt{\mathcal{Q}}$, the previous argument applies also to the case $\bar\psi\psi=0$ $(K=0)$ giving rise to the condition $\mathcal{Q}={\rm const.}$.

At this point, the last step would be to verify that the relation $\mathcal{Q}\left(\tau\right)=-2 \left( \frac{\dot{a}\dot{b}}{ab} + \frac{\dot{a}\dot{c}}{ac} +\frac{\dot{b}\dot{c}}{bc} \right)$ is preserved over time. This final requirement is seen to select suitable relationships between the admissible integration constants appearing in the found solutions. 

In order to show how the above outlined procedure works, we consider the model $f(\mathcal{Q})=\alpha\/\mathcal{Q}^n$ (with $n$ natural odd number for brevity) as an example. In this case, from eq. \eqref{4.13} we deduce the relation
\be\la{4.18}
\mathcal{Q}=\left[\frac{H}{\left(1-2n\right)\alpha}\right]^{\frac{1}{n}}\left(\frac{1}{\tau}\right)^{\frac{1}{n}}
\ee
and thus
\be\la{4.19}
f(\mathcal{Q}(\tau))=\frac{H}{\left(1-2n\right)\tau} \quad {\rm and} \quad 
f'(\mathcal{Q}(\tau))=\alpha\/n\left[\frac{H}{\left(1-2n\right)\alpha}\right]^{\frac{n-1}{n}}\left(\frac{1}{\tau}\right)^{\frac{n-1}{n}}
\ee
where we have set $H:=mK$ for simplicity. If we require expansion in all three spatial directions, we must impose the condition $\frac{H}{\left(1-2n\right)\alpha}<0$. Under the same hypothesis, if the exponent n had been even, we would have had to define the quantity \eqref{4.18} with the opposite sign, demanding $\frac{H}{\left(1-2n\right)\alpha}>0$ as well. 

Inserting eqs. \eqref{4.18} and \eqref{4.19} into eq. \eqref{4.16} and after a first integration step, we end up with the differential equation
\be\la{4.20}
\dot\tau = \frac{1}{\alpha\/n}\left(\frac{\alpha\left(1-2n\right)\tau}{H}\right)^{\frac{n-1}{n}}\left(-\frac{3H t}{4\left(1-2n\right)}+H_0\right)
\ee
where $H_0$ is an integration constant. Eq. \eqref{4.20} admits exact solutions of the form
\be\la{4.21}
\tau(t)=\left[-\frac{\left(\alpha\left(1-2n\right)\right)^{-\frac{1}{n}}}{n^2}\left(H_0\left(2n-1\right)H^{\frac{1-n}{n}}+\frac{3}{8}H^{\frac{1}{n}}t\right)t + H_1\right]^n
\ee
where $H_1$ is again an integration constant. As already mentioned, the integration constants $H_0$, $H_1$, $d_i$ and $g_i$ have to be chosen in such a way that the relation $\mathcal{Q}\left(\tau\right)=-2 \left( \frac{\dot{a}\dot{b}}{ab} + \frac{\dot{a}\dot{c}}{ac} +\frac{\dot{b}\dot{c}}{bc} \right)$ is preserved over time. For instance, in this specific case, it is easily seen that the condition $H_0=0$ and $H_1=0$ is only compatible with the choice $d_1=g_1=0$, namely with a totally isotropic universe; instead, the values $H_0=0$ and $H_1=\frac{2g_1^2}{3n^2\alpha^2}\left(\frac{H}{\alpha\left(1-2n\right)}\right)^{\frac{1-2n}{n}}$ are compatible with a partially isotropic universe $a=b$ ($d_1=d_2=0$).

As a last remark, it is worth noticing that the following identity holds
\be\la{4.22}
f'\left(\mathcal{Q}\left(\tau\right)\right)\tau = \alpha\/n\left[\frac{H}{\left(1-2n\right)\alpha}\right]^{\frac{n-1}{n}}\tau^{\frac{1}{n}}
\ee
and therefore the scale factors \eqref{4.17} necessarily isotropize at late cosmological time. Moreover, by suitably choosing the value of the exponent $n$, still at late time we may have accelerated expansion for all the scale factors. This result is confirmed solving for the scale factors. Indeed, setting
\be\label{conti2}
A^2=\frac{3}{8n^2}\left[\frac{H}{\alpha(2n-1)}\right]^{\frac{1}{n}} , \quad   B^2=\frac{2g_1^2}{3n^2\alpha}\left[\frac{H}{\alpha(2n-1)}\right]^{\frac{1-2n}{n}},\quad
C=\alpha\/n\left[\frac{H}{\alpha(1-2n)}\right]^{\frac{n-1}{n}}
\ee
from eqs. \eqref{4.17}, \eqref{4.21} and \eqref{4.22} we have
\begin{subequations}\label{conti7}
\be\label{conti7a}
a(t)=b(t)= \left[A^2t^2-B^2\right]^{\frac{n}{3}}e^{\frac{g_2}{3}}\left[\frac{t-(B/A)}{t+(B/A)}\right]^{\left(\frac{g_1}{6ABC}\right)}
\ee
\be\label{conti7b}
c(t)=\left[A^2t^2-B^2\right]^{\frac{n}{3}}e^{\frac{-2g_2}{3}}\left[\frac{t-(B/A)}{t+(B/A)}\right]^{\left(\frac{-g_1}{3ABC}\right)}
\ee
\end{subequations}

\begin{figure}
    \centering
    \includegraphics[width=0.7\textwidth]{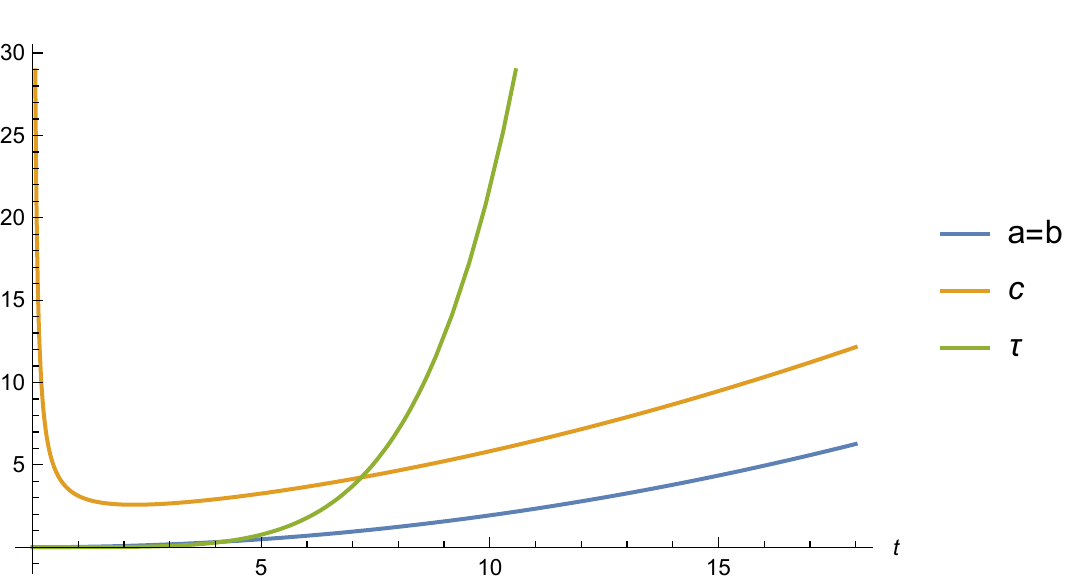}
    \caption{Plot of the scale factors \eqref{conti7} for $H=3$, $\alpha=6$, $n=3$, $g_1=1$, $g_2=0$. the plot has been translated in such a way that the scale factors are zero at $t=0$.}
    \label{fig:my_label}
\end{figure}

\section{Conclusions}\label{sec.V}

In this paper, we have presented a tetrad--affine approach to $f(\mathcal{Q})$ gravity coupled to spinor fields of spin-$\frac{1}{2}$. The proposed formulation relies on the adoption of unusual pairs of dynamical variables $(e^\mu_i,\Gamma_{ij}^{\;\;\;h})$, consisting of a tetrad field and an affine connection. This choice has been motivated at first by the necessity to have more treatable field equations, but then it has revealed some interesting features of $f(\mathcal{Q})$ gravity.

The use of the affine connection $\Gamma_{ij}^{\;\;\;h}$, instead of the more usual spin connection $\omega_{i\;\;\;\nu}^{\;\;\mu}$, implies the appearance of additional terms in the energy-momentum tensor which modify the standard Dirac tensor. These additional terms involve the covariant derivatives of the spin density and can be elaborated by using the conservation law for the spin, directly deduced by the Dirac equations. As with metric theories with torsion, it has been proved that the spin conservation law ensures that the antisymmetric part of the Einstein-like equations is satisfied. 

After proving consistency between the derived field equations, we have analyzed Bianchi type-I cosmologies in the context of $f(\mathcal{Q})$ theory. We have shown that, as in the case of $f(R)$ theories with torsion, the off-diagonal part of the gravitational field equations imposes restrictions on both the geometry and the spinor field.

In order for the constraints mentioned above to be satisfied, three different scenarios are possible: i) an isotropic spacetime where no further restrictions are imposed on the spinor field; ii) an anisotropic spacetime where both the spin four-vector and the scalar $\bar\psi\psi$ are zero, and thus a spacetime where the spinor field does not contribute to cosmological dynamics; iii) a universe where two scale factors are identical, and only one spatial component of the spin four-vector does not vanish. 

Finally, we have proposed a general procedure to solve the resulting field equations, reducing the dynamical problem to a single differential equation for the spatial volume $\tau$. To show how the given procedure works, we have considered gravitational Lagrangians of the kind $f(\mathcal{Q})=\alpha\/\mathcal{Q}^n$. The corresponding dynamical problem has been analytically solved, showing that such models can give rise to initially anisotropic universes that isotropize with accelerated expansion.



\end{document}